\documentclass[12pt,preprint]{aastex}
\usepackage{natbib}
\newcommand{\Htwo}{\mbox{H$_{2}$}}
\newcommand{\Htwoo}{\mbox{H$_{2}$O}}
\newcommand{\Htwoco}{\mbox{H$_{2}$CO}}

\newcommand{\cotwo}{\mbox{CO$_{2}$}}
\newcommand{\cotwop}{\mbox{CO$_{2}\!^{+}$}}

\newcommand{\xsig}{$X\,^1\Sigma ^{+}$}

\newcommand{\atpi}{$a\,^3\Pi$}
\newcommand{\bsig}{$B\,^1\Sigma ^{+}$}
\newcommand{\csig}{$C\,^1\Sigma ^{+}$}
\newcommand{\epi}{$E\,^1\Pi~$}
\newcommand{\kms}{km~s$^{-1}$}

\newcommand{\iue}{\textit{IUE\/}}
\newcommand{\hst}{\textit{HST\/}}
\newcommand{\fuse}{\emph{FUSE\/}}

\shorttitle{Formaldehyde in Comets}
\shortauthors{Feldman, Lupu, McCandliss, and Weaver}

\begin{document}

\title{The Far Ultraviolet Spectral Signatures of Formaldehyde 
and Carbon Dioxide in Comets}

% USE FULL NAME
\author{Paul D. Feldman, Roxana E. Lupu, and Stephan R. McCandliss}

\affil{Department of Physics and Astronomy, The Johns Hopkins University\\ 
3400 N. Charles Street, Baltimore, Maryland 21218}
\email{pdf@pha.jhu.edu}

\and
\author{Harold A. Weaver}

\affil{Space Department,
Johns Hopkins University Applied Physics \mbox{Laboratory,}
11100 Johns Hopkins Road,
Laurel, MD 20723-6099}

%%%% SEE BELOW FOR HOW TO ADD AUTHORS WITH ALT AFFILIATIONS

%\pagestyle{myheadings}
%\markright{Revised \today}

\begin{abstract}

Observations of four comets made with the {\it Far Ultraviolet
Spectroscopic Explorer} show the rotational envelope of the (0,0) band
of the CO Hopfield-Birge system (\csig -- \xsig) at 1088 \AA\ to consist
of both ``cold'' and ``hot'' components, the ``cold'' component
accounting for $\sim$75\% of the flux and with a rotational temperature
in the range 55--75~K.  We identify the ``hot'' component as coming
from the dissociation of \cotwo\ into rotationally ``hot'' CO, with
electron impact dissociation probably dominant over photodissociation near the nucleus.  An additional weak, broad satellite band is seen
centered near the position of the P(40) line that we attribute to CO
fluorescence from a non-thermal high $J$ rotational population produced
by photodissociation of formaldehyde into CO and \Htwo.  This process
also leaves the \Htwo\ preferentially populated in excited vibrational
levels which are identified by fluorescent \Htwo\ lines in the spectrum
excited by solar \ion{O}{6} $\lambda$1031.9 and solar Lyman-$\alpha$.  
The amount of \Htwo\ produced by \Htwoco\ dissociation is comparable
to the amount produced by photodissociation of \Htwoo.
Electron impact excitation of CO, rather than resonance fluorescence,
appears to be the primary source of the observed \bsig -- \xsig (0,0)
band at 1151~\AA.

\end{abstract}

%\keywords{topics alphabetically --- subsubtopic: topic --- subtopic, subtopic}
\keywords{comets: individual (C1999 T1, C/2001 A2, C/2000 WM1, C/2001 Q4) --- 
ultraviolet: solar system}

\newpage
\section{INTRODUCTION}

In previous papers \citep{Feldman:2002,Weaver:2002,Feldman:2005}, we
reported on the spectra of four comets observed by the {\it Far
Ultraviolet Spectroscopic Explorer} (\fuse).  Launched in June 1999,
\fuse\ operated through October~2007 providing an orbiting capability
with a spectral resolution better than 0.4 \AA\ in the wavelength range
905--1187~\AA\ together with very high sensitivity to weak emissions
making possible both the search for minor coma species and extensive
temperature and density diagnostics for the dominant species.  Our
initial reports on the three comets observed in 2001 focused on three
Hopfield-Birge band systems of CO not previously observed in comets,
the \bsig -- \xsig (0,0) band at 1150.5 \AA, the \csig -- \xsig (0,0)
band at 1087.9~\AA, and the weak \epi -- \xsig (0,0) band at
1076.1~\AA, three lines of H$_2$ fluorescently pumped by solar
Lyman-$\beta$ radiation, and upper limits to emission from \ion{O}{6}
and \ion{Ar}{1}.  The rotational envelopes of the CO bands are resolved
and appear to consist of both cold and warm components, the cold
component, with a rotational temperature in the range 55--75~K,
accounting for over 75\% of the flux and presumably due to native CO
released by the nucleus.  The warm ($\sim$500~K) component suggested a
\cotwo\ photodissociation source but this was considered unlikely
because the low rate for this process in the solar radiation field
would have required a very high relative abundance of \cotwo.

\ion{O}{6} emission at 1031.9 and 1037.6~\AA\ was searched for but only
the 1031.9~\AA\ line was marginally detected in C/2000 WM1 at about the
level predicted by a model used to explain the X-ray emission from
comets as being produced by charge exchange between solar wind ions and
cometary neutrals \citep{Kharchenko:2001}.  The higher signal/noise
ratio in the spectrum of comet C/2001 Q4, observed in April 2004,
enabled an accurate determination of the wavelength of this feature
which suggested that it was the \Htwo\ (1,1)~Q(3) Werner ($C\,^1\Pi_u -
X\,^1\Sigma_g^+$) line, which can be fluorescently pumped by either
solar \ion{N}{3}~$\lambda$989.8 or \ion{O}{6}~$\lambda$1031.9.  The
latter requires the \Htwo\ to be in the first vibrationally excited
state \citep{Liu:1996}.  The Q(3) lines of the (1,3) band at
1119.1~\AA\ and the (1,4) band at 1163.8~\AA\ are also detected as
would be expected for fluorescent excitation.  However, the ground
state \Htwo\ abundance in the coma, estimated from the solar
Lyman-$\beta$ pumped fluorescence of the Lyman ($B\,^1\Sigma_u^+ -
X\,^1\Sigma_g^+$) (6,1) P1 line, is insufficient for excitation by
solar \ion{N}{3} (a factor of 10 weaker than \ion{O}{6}), which implies
the formation of \Htwo\ in the $v = 1$ state and excitation by solar
\ion{O}{6}.

The molecular channel of the photodissociation of formaldehyde (i.e.,
the channel leading to \Htwo\ and CO products) has been well studied
both experimentally and theoretically.  Its products are characterized
by a non-thermal rotationally ``hot'' CO molecule and a vibrationally
excited \Htwo\ molecule \citep{vanZee:1993}.  The CO rotational
distribution can be represented by a gaussian centered near $J = 40$
with a half-width at half maximum in $J$ of about 20.  Fluorescence
from such levels in the P-branch (the R-branch is shortward of the
detector wavelength cut-off) is clearly seen in two of the comets
observed by \fuse\ and we propose that this feature, together with the
vibrationally excited \Htwo, constitute an unambiguous signature of
\Htwoco\ in a cometary coma.  Formaldehyde has been observed in these
comets in both the infrared and millimeter spectral regions
\citep{Biver:2006, Milam:2006, Gibb:2007}.  The capability of \fuse\ to
identify \Htwoco\ in a cometary coma provides a new means to determine
its production rate relative to that of water and to compare these
results with ground-based observations in the infrared and millimeter
regions of the spectrum.

Since the thermal ``hot'' component of the observed CO emission is not
produced by \Htwoco\ dissociation, we return to \cotwo\ as a possible
parent, particularly as the rotational temperature is what would be
expected based solely on angular momentum conservation arguments
\citep{Mumma:1975}.  We suggest that within the \fuse\ field-of-view,
photoelectron impact dissociation of \cotwo\ has a comparable rate to
photodissociation and we base this conclusion on estimates of the mean
photoelectron flux derived from the other prominent CO band observed in
these spectra, the $B - X$~(0,0) band at 1151~\AA.  We thus have 
unique diagnostics in which the shapes of these molecular bands can be
used to infer the column abundance of CO from both its native source as
well as from two of its direct molecular parents, together with the
photoelectron contributions to excitation and dissociation in the
inner coma.

\section{OBSERVATIONS}

Three long period comets were observed during 2001, C/1999 T1
(McNaught-Hartley), C/2001 A2 (LINEAR), and C/2000 WM1 (LINEAR), with
C/2001 A2 being the most active of the three and showing the brightest
far-ultraviolet emissions.  In December 2001, two of the reaction
wheels on the spacecraft failed resulting in a shutdown of observatory
operations until a new pointing control program could be devised and
implemented a few months later.  However, no moving target observations
were attempted until early in 2004 when comet C/2001 Q4 (NEAT) was
observed.  The observation parameters for all four comets are
summarized in Table~\ref{tab1}.

\placetable{tab1}

Observations of each comet were made over contiguous spacecraft orbits
during which the center of the $30'' \times 30''$ spectrograph aperture
was placed at the ephemeris position of the comet and tracked at the
predicted sky rate.  Due to the extended, though not uniform, nature of
the cometary emissions within the aperture, the effective spectral
resolution is $\sim$0.25~\AA.  For each comet, all of the orbits of
data were co-added and the extracted fluxes were converted to average
brightness (in rayleighs) in the $30'' \times 30''$ aperture.  In
addition, the data were separately co-added for only the time during
which the spacecraft was in Earth shadow.  These ``night-only'' spectra
make it possible to differentiate cometary emissions of \ion{H}{1},
\ion{O}{1}, \ion{N}{1}, and, in second order, \ion{He}{1}, from the
same emissions produced in the daytime terrestrial atmosphere  \citep{Feldman:2001}.  For the CO and \Htwo\ emissions this is not a
problem and the entire data set for each comet was used.

All of the data were reprocessed using the final version of the
\fuse\ pipeline, {\it CalFUSE 3.2} \citep{Dixon:2007}.  Among the
enhancements incorporated in this version were temperature-dependent
wavelength corrections, time-dependent corrections to the detector
effective areas, and a more complete evaluation of detector
background.  The latter was particularly significant in allowing
reliable extractions of the fluxes of weak emission features.  The
overall absolute flux calibration, based on standard stars and adjusted
for the date of observation, is better than 5\%.  Table~\ref{tab1} also
lists the extracted brightnesses of the CO emission features discussed
below.

\section{SPECTRA}
\subsection{Data}
 
\placefigure{cx1}
\placetable{tabn}

The CO $C-X$ (0,0) band, from the total exposure, is shown for each
comet in Figure~\ref{cx1}.  The P and R branches appear well separated,
characteristic of a rotational temperature of 55--75~K.  There also
appear broad wings to the band indicative of a rotational temperature
in the range of 500 to 600 K.  The figures also show a model fit,
assuming solar resonance fluorescence, obtained using the procedure
described by \citet{Feldman:2002}.  The derived column densities and
rotational temperatures are given in Table~\ref{tabn}.  In the two most
active comets, C/2001 A2 and C/2001 Q4, there is also an additional
broad feature centered at $\sim$1089.4~\AA, that we identify as
P-branch lines from high-lying $J$ states.  The position of the P(40)
line, calculated using the rotational constants of 
\citet{Eidelsberg:1991}, is also shown in Fig.~\ref{cx1}.  We note that
the \citeauthor{Eidelsberg:1991} wavelengths have been verified
experimentally only up to $J = 25$.  The corresponding high $J$ R-lines
are off the short wavelength edge of  the LiF2a detector.  A ``hot'' 
thermal rotational distribution arises in molecular
photodissociation in order to conserve angular momentum when the
product molecule has a much smaller moment-of-inertia than the parent
molecule as is the case for \cotwo\ \citep{Mumma:1975}.  The
non-thermal component is identified with CO as a dissociation product
of \Htwoco, and the column abundance of this component, also given in
Table~\ref{tabn}, is derived assuming an equal contribution from the
unobserved R-branch.  In this case, an optically thin band fluorescence
efficiency is used as the CO population is spread over many rotational 
levels and the absorption is not likely to be optically thick.

\subsection{Photodissociation of Formaldehyde}

As noted above, the photodissociation of formaldehyde has been
intensively studied, both experimentally and theoretically.  The lowest
dissociation threshold ($\sim$3600~\AA) leads to \Htwo\ + CO (the
``molecular'' channel) while $\sim$2700 cm$^{-1}$ higher
($\sim$3300~\AA), H + HCO (the ``radical'' channel) can also be
produced \citep{vanZee:1993}.  The nature of the \Htwo\ and CO products
differs markedly above and below this second energy threshold.  Consider first
the region below the radical channel threshold.  \citet{Bamford:1985}
and \citet{Debarre:1985} have used laser-induced fluorescence
spectroscopy and coherent anti-Stokes Raman scattering to
experimentally study the product state populations of CO and \Htwo.
\citeauthor{Bamford:1985} found a nonthermal highly excited
CO rotational population with a Gaussian distribution centered at
$J=40$ and a half-width at half maximum of 20 $J$ units, but with little
vibrational excitation.  Later work \citep[e.g.,][]{vanZee:1993} showed
that the shape was dependent on the energy of the absorbed photon.  It
should be noted that the absorption of \Htwoco\ in this region is in
discrete bands \citep{Rogers:1990}.  For \Htwo,
\citeauthor{Debarre:1985} found that the primary products are the ortho
($J = 3$ and 5) states of $v = 1$.  For photolysis at 3390~\AA\ they
found the fractional population of the $v = 1$, $J = 3$ level to be
11\%.  \citet{Butenhoff:1990} extended this work to additional
wavelengths and found a peak \Htwo\ population in $v=1, J=3$.

Above the radical channel threshold, \Htwo\ and CO are found to have
markedly different populations with the non-translational energy going
principally to vibrational excitation of \Htwo\ while the CO rotational
state population shifts to lower $J$ states \citep{vanZee:1993}.
\citet{Zhang:2005} demonstrated that the population of vibrationally
excited \Htwo, up to $v=8$ or 9, increased as the energy of the
exciting photon increased with a corresponding increase in the low-$J$
state population of CO such that the non-translational energy was
conserved.  However, the peak at $J=40$ remained and the low-$J$
population was still highly non-thermal.  \citet{Chambreau:2006} showed
that the excited vibrational levels of \Htwo\ also had peak
rotational populations at $J = 5$~or 7, depending on the photolysis
energy.

In the solar radiation field in the coma of a comet, the \Htwo\ and
CO state populations derived from \Htwoco\ dissociation will reflect a
superposition of populations, weighted by the solar flux and absorption
cross section \citep{Huebner:1992}.  Such a calculation is beyond the
scope of the present paper.  The broad, weak feature seen longward of
the $C-X$ (0,0) band in comets C/2001~A2 and C/2001~Q4 in
Fig.~\ref{cx1} is centered near the position of the P(40) line of this
band, and the non-thermal high-$J$ population of this feature is an
unambiguous fingerprint of CO produced from
\Htwoco\ photodissociation.  Although the spectrum of C/2000~WM1 is
significantly noisier, the data also suggest enhanced emission on the
red wing of the band in this comet.  Further evidence for this source
is the simultaneous detection of fluorescence from vibrationally
excited \Htwo\ in this comet \citep{Liu:2007} as well as in comets
C/2001~A2 and C/2001~Q4.

\subsection{Fluorescence of Vibrationally Excited H$_2$ \label{vib}}

\placefigure{h2}

In Figure~\ref{h2} the region of the \ion{O}{6} doublet in comet
C/2001~Q4 is shown, and the \Htwo\ (1,1)~Q(3) line is identified,
separated by 0.06 \AA\ from the position of the \ion{O}{6} line at
1031.92~\AA.  The uncertainty in the \fuse\ wavelength scale is
$\sim$0.01~\AA.  The same figure also shows the corresponding
(1,3)~Q(3) and (1,4)~Q(3) lines at 1119.09 and 1163.79~\AA,
respectively \citep{Liu:1996}.  Note that there is no emission at the
expected position of the \ion{O}{6} line at 1037.62~\AA.  These lines
are quite weak and with the exception of a 3-$\sigma$ detection in
comet C/2000 WM1 \citep{Liu:2007} do not appear in the spectra of the
two other comets observed by \fuse.

The absence of these lines in the other two comets is the result of the
heliocentric Doppler shift of the exciting solar line, the Swings
effect \citep{Feldman:2004}, rather than the lack of \Htwo\ in the
coma.  This can be illustrated with the solar
\ion{O}{6}~$\lambda$1031.9 line profile.  \citet{Doschek:2004} have
used the recent solar spectra taken by the SUMER instrument on {\it
SOHO} with an instrumental resolution of $\sim$0.1~\AA, to determine
the width of this line under quiet sun conditions to be 0.202~\AA.
This result is from an observation of a small part of the solar disk
near the center, while for the calculation of cometary excitation rates
the whole disk flux is more useful.  Fortuitously, we are able to take
advantage of a minor operational problem of \fuse\ \citep{Sahnow:2000},
a sensitivity at certain orientations on the sky of the silicon carbide
(SiC) channels to scattered solar radiation, to determine the whole
disk line shape.  The observation of comet C/2001~Q4 resulted in such 
an orientation and the SiC1b and SiC1a channels (905--992 and 
1005--1090~\AA, respectively) measured a large scattered solar flux in
all three apertures.  Both the high (HIRS, $1.\!''5 \times 20''$) and
medium resolution (MDRS, $4'' \times 20''$) apertures obtained good S/N
spectra of the \ion{O}{6} lines at instrumental resolutions of 0.035
and 0.075~\AA, respectively.  From both of these channels we derived a
width of 0.200~\AA\ for the solar \ion{O}{6}~$\lambda$1031.9 line, in
excellent agreement with the result of \citeauthor{Doschek:2004}.  The
MDRS profile is shown in Figure~\ref{solar}, together with the Doppler
shifted positions of the \Htwo\ (1,1)~Q(3) line for each comet.  Note
that the profile shown in the figure suggests that the solar line is
not adequately represented by a single Gaussian.  Clearly, the
detection of the (1,$v''$) lines in comet C/2001~Q4 was favored by its
negative heliocentric velocity and by the relatively high activity level
of this comet at the time of observation.
 
\placefigure{solar}

\citet{Wolven:1997} noted that solar Lyman-$\alpha$ (1216.57 \AA)
could pump vibrationally excited \Htwo\ in the $v=2$ level via the
Lyman band (1,2)~P(5) and R(6) lines, and used this mechanism to
identify fluorescence from the upper level in {\it Hubble Space
Telescope} spectra of Jupiter following the impact of comet
Shoemaker-Levy 9 in July 1994.  The same mechanism has recently been
found to operate in planetary nebulae where the \fuse\ spectra show
the (1,1)~P(5) and R(6) lines at 1161.82 and 1161.95 \AA, respectively
\citep{Lupu:2006}.  One of the previously unidentified lines reported
in \fuse\ spectra of comets at 1161.83~\AA\ was identified by
\citet{Liu:2007} as a blend of these two \Htwo\ lines (see
Fig.~\ref{h2}).  The vibrationally excited \Htwo\ would most likely
come from \Htwoco\ rather than \Htwoo\ photodissocation as proposed by
\citeauthor{Liu:2007}  This feature has roughly the same brightness in
comets C/2001~A2 and C/2001~Q4, and we note that the brightnesses of
the CO~P(40) feature in these two comets are also
comparable.  From the branching ratios given by \citeauthor{Lupu:2006},
we would also expect to see the (1,1)~R(3) line at 1148.70~\AA\ and the
(1,1)~P(8) line at 1183.31~\AA\ from the same excited $J$~levels, and
indeed we do.  The latter is considerably weaker than the former due to
the lower population of the even-$J$ (para) states as expected.

\citeauthor{Liu:2007} also identified several other cometary lines as
being pumped by solar Lyman-$\alpha$ from higher vibrational and
rotational levels of \Htwo.  Almost all of the strong unidentified
features tabulated by \citet{Feldman:2005} are amongst them.
Our analysis supports the conclusion of \citeauthor{Liu:2007} that the
vast majority of emission features in the \fuse\ spectra of comets can
be attributed to \Htwo. However, unlike \citeauthor{Liu:2007}, who
proposed that the excited vibrational and rotational levels of
\Htwo\ arise from the photodissociation of \Htwoo, we suggest instead
that these excited \Htwo\ states are produced during the
photodissociation of \Htwoco\ because their presence in the cometary
coma is consistent with the \Htwo\ level populations measured during
\Htwoco\ photodissociation in the laboratory by \citet{Zhang:2005} and
\citet{Chambreau:2006}.

\subsection{Electron Impact Excitation of CO and Dissociation of \cotwo
\label{electron}}

\placefigure{bx1}

The CO $B-X$~(0,0) band at 1151~\AA\ is weaker than the $C-X$~(0,0)
band by a factor of 3 to 4 and thus recorded with poorer signal/noise
in all four comets as shown in Figure~\ref{bx1}.  It is surprising that
this band is observed at all since its fluorescence efficiency is 44
times lower than that of the $C-X$~(0,0) band (the predicted spectra
using the CO column densities given in Table~\ref{tabn} are also shown
in the figure).  There must be an additional source of excitation, and
since the band shape is consistent with the low temperature CO
component, that can only be electron impact excitation of CO.
Photoelectrons in the inner comae of comets have been considered by
several authors \citep[e.g.,][]{Ashihara:1978,Koros:1987}, and were
detected in the {\it Vega} fly-by of 1P/Halley \citep{Gringauz:1986},
but their role in excitation, dissociation and ionization is difficult
to quantify.

An equivalent to the fluorescence efficiency, or g-factor, may be
defined by an electron excitation rate per molecule per second, $g'$:

\[ g' = \int\int F(E) \sigma(E) dE d\Omega \]

\noindent
where $F(E)$ is the photoelectron flux in 
electrons~s$^{-1}$cm$^{-2}$sr$^{-1}$eV$^{-1}$ and $\sigma(E)$ is
the cross section in cm$^2$.  In general, the photoelectron flux is
a function of cometocentric distance and is anisotropic so that $g'$
will vary over the spectrograph field-of-view.

\placefigure{cross}

Rather than attempt to use modeled photoelectron fluxes, we
instead use the observed CO $B-X$~(0,0) band brightnesses to derive
an effective $g'$ over the \fuse\ aperture but then use the modeled
electron energy distribution to determine the relative excitation
rates for other processes.  In particular, we are interested in the
electron impact dissociation rate of \cotwo.  Relevant cross sections
in analytic form for electron impact on CO and \cotwo\ are available
from the compilation of \citet{Shirai:2001}.  Figure~\ref{cross} shows
the cross sections for electron impact excitation on CO of the two
bands under discussion and that for electron impact dissociation of
\cotwo\ into the metastable \atpi\ state of CO.  A typical photoelectron
energy dependence, from \citet{Koros:1987}, is also shown.  The \atpi\
state is the upper level of the CO Cameron bands, whose excitation in
comets has been discussed by \citet{Weaver:1994}.  The total \cotwo\
dissociation cross section has been studied by \citet{Cosby:1992}.
These authors note that the branching ratio into the \atpi\ state of CO
is 27\% of the total electron dissociation rate, the same as for
solar photodissociation \citep{Huebner:1992}.

\placetable{tabg}

The derived values of $g'$, reduced to 1 AU, are given in
Table~\ref{tabg}.  For comparison, the optically thin solar
fluorescence efficiencies are also given.  Figure~\ref{cross} shows
that the steep increase of photoelectron flux towards lower energies
favors the $B-X$ transitions over the $C-X$.  Even so, the electron
impact contribution to $C$ state excitation will reduce the derived CO
column abundances by 5--15\% and this is reflected in the values given
in Table~\ref{tabn}.  These numbers remain uncertain as both the
photoelectron energy distribution and flux will vary with increasing
column density towards the nucleus.  This will also affect the ratio of
total \cotwo\ dissociation to that fraction leading to the \atpi\ state
of CO.  The $g'$ values for these processes, for the photoelectron
model adopted, are also given in Table~\ref{tabg}.  For comparison with
photodissociation at 1~AU, the solar maximum value given by
\citet{Huebner:1992} is $2.56 \times 10^{-6}$~molecule$^{-1}$~s$^{-1}$,
so electron impact dissociation and photodissociation are comparable.

Evidence for a significant role of photoelectrons in the inner coma
has often been noted in the literature.  Two particular cases in which
photodissociation and photoionization were inadequate to account for
the abundance of daughter products are the production of atomic carbon
in the metastable $^1$D~state from CO \citep{Tozzi:1998} and the high
concentration of \cotwop\ close to the nucleus seen in early
\iue\ spectra \citep{Festou:1982}.  The \fuse\ spectra provide
additional evidence in the strong \ion{O}{1}~$^1$D~--~$^1$D$^o$~line
at 1152.16~\AA, seen in Fig.~\ref{bx1}, which also cannot be accounted for by
solar resonance scattering alone.  The lower state of this transition,
the metastable ~$^1$D~state, is an abundant product of photodissociation
of \Htwoo, OH, and CO, and is the upper level of the oxygen ``red'' lines
at 6300~and~6364~\AA, seen in many comets and often used as a surrogate
for the water production rate \citep{Feldman:2004}.

\section{DISCUSSION}

\subsection{Modeling \label{model}}

The column density of the non-thermal component of CO was modeled assuming
an \Htwoco\ parent using the vectorial model of \citet{Festou:1981}
with solar photodestruction rates given by \citet{Huebner:1992}.  
\citet{Meier:1993} have also evaluated the photodissociation lifetimes
(\Htwoco\ has two other dissociation channels, H + HCO, and H + H + CO)
and find the rate into \Htwo\ + CO to be 25\% lower than that of
\citeauthor{Huebner:1992} while the total rate is 7\% lower.
\citet{Bockelee-Morvan:1992} find a total photodissociation lifetime in
close agreement with that of \citeauthor{Huebner:1992}  The model
assumes a nucleus source for the \Htwoco.  Even if a distributed source
were present (see Section~\ref{dist} below), the size of the
\fuse\ aperture translates to a few thousand km projected on the sky
and such a source can be ignored for our purposes.

To determine production rates relative to water, we use values of
$Q_{\rm H_2O}$ taken from the recent literature; values derived from
SOHO/SWAN measurements of cometary Lyman-$\alpha$ emission by 
\citet{Combi:2008}, and those derived from ODIN submillimeter observations
of \Htwoo\ directly by \citet{Biver:2007}.  Where they overlap, these two
sets of values are in agreement to $\sim$20--25\%.  We adopt the values
from \citeauthor{Combi:2008} because of their finer sampling grid, allowing
us to use data from the same day as the \fuse\ observations.  We use
the deconvolved model results from \citeauthor{Combi:2008} rather than the
daily averages given.  For C/2001 Q4 we use a value derived from 
near-simultaneous \hst\ observations of OH emission, which is close
to that reported by \citeauthor{Biver:2007}

For C/2001 A2 we note a significant difference from the value of
$Q_{\rm H_2O}$ cited in the initial reports of the \fuse\ observations
by \citet{Feldman:2002} and \citet{Weaver:2002}.  The earlier estimate
was based on a visual outburst two days prior to the observation and a
modeled production rate adjusted to match the derived \Htwo\ column
density.  The latter was based on fluorescence efficiencies derived by
\citet{Krasnopolsky:2001a}.  With the availability of detailed solar
Lyman-$\beta$ fluxes and line shapes from SOHO/SUMER
\citep{Lemaire:2002}, the heliocentric velocity dependent g-factors
were recalculated and the \Htwo\ column density revised downward by a
factor of about three, which is consistent with the water production
rate for the date given by \citeauthor{Combi:2008}

\placetable{tabq}

The results are given in Table~\ref{tabq}.  For the comets in
which the non-thermal CO is detected, the production rates of
\Htwoco\ relative to that for \Htwoo\ are generally in the range of
values found from infrared and millimeter observations.
\citet{Biver:2006} reported on four comets observed at millimeter
wavelengths including the three comets observed by \fuse\ at comparable
epochs in 2001.  For comets C/2001~A2 and C/2000~WM1, our relative
production rates are comparable to those of \citeauthor{Biver:2006} based
on their model that assumes a distributed source with an effective parent
lifetime about a factor of 2 longer than the true photochemical
lifetime.  With an assumed \Htwoco\ parent, \citeauthor{Biver:2006} derive
relative production rates $2-4$ times smaller.  

Millimeter observations of comet C/2001~Q4 were made in mid-May~2004 by
\citet{Milam:2006}.  Their results for \Htwoco, again assuming a
distributed grain source, are in close agreement with the
\fuse\ value.  While infrared bands of \Htwoco\ near $3.6~\mu$m have
been detected in a number of recent comets, the only data available for
comparison with the \fuse\ observations is of C/2001~A2
\citep{Gibb:2007}.  \citeauthor{Gibb:2007}, observing 2 and 3 days
prior to the \fuse\ observation, found a low value of 0.24\% for the
ratio of the production rate of \Htwoco\ relative to \Htwoo\ on July~9.5
followed by a decrease of a factor of 4 one day later. In contrast, we
derive a much higher value for this ratio on July~12.5.  The apparent
variability of \Htwoco\ is discussed below in Section~\ref{caveat}.

Modeling the \Htwo\ brightness is not as straightforward.  The \Htwo\
molecules initially carry away most of the excess energy of
dissociation as translational energy ($\sim$11 \kms), but they are
rapidly thermalized in the inner coma \citep{Combi:2004}.  The excited
state populations will also vary with incident wavelength and be
altered by collisions, so the actual number of \Htwo\ molecules in the
$v = 1, J = 3$ level produced by solar photodissociation will have a
large uncertainty.  Nevertheless, as a consistency check we can
evaluate the number of such molecules in the field-of-view needed to
produce the observed 0.13 rayleigh brightness of the (1,1) Q(3) line at
1031.86 \AA\ observed in comet C/2001~Q4.  The fluorescence efficiency,
or g-factor, at the peak of the solar \ion{O}{6} line is calculated
using the solar minimum irradiance recommended by \citet{Warren:2005}
and the oscillator strengths and branching ratios taken from the
compilation of \citet{Abgrall:1993b}.  The derived g-factor is $3.8
\times 10^{-7}$ s$^{-1}$~molecule$^{-1}$ at 1 AU, so that a mean column
density of $3.6 \times 10^{11}$ cm$^{-2}$ is sufficient to produce the
observed emission rate.  From Table~\ref{tabn}, the column density of
non-thermal CO in this comet is $4.1 \times 10^{12}$ cm$^{-2}$, and
78\% of those are accompanied by \Htwo\ rather than by two H atoms
\citep{Huebner:1992}.  Thus 11\% of the total \Htwo\ needs to be in $v
= 1$, $J = 3$, which is plausible based on the laboratory data of
\citet{Zhang:2005} and \citet{Chambreau:2006}.  Given all of the
uncertainties, this is good agreement and supports our proposal that
\Htwoco\ is the source of both the non-thermal CO and the vibrationally
excited \Htwo\ detected by \fuse.

The \Htwo\ vibrational population may also be examined with the
identification of solar Lyman-$\alpha$ pumping of lines in the Lyman
(1,2) band as described in Section~\ref{vib}.  The solar Lyman-$\alpha$
spectral flux is roughly two orders of magnitude larger than that of
\ion{O}{6}~$\lambda$1031.9 resulting in comparably larger fluorescence
efficiencies for these lines.  Using solar minimum
Lyman-$\alpha$ fluxes from \citet{Lemaire:2002} and molecular data for
the Lyman bands from \citet{Abgrall:1993a}, the g-factors for the
(1,1)~P(5) and R(6) lines can be calculated.  The relative population
of the $v = 2$, $J = 5$, needed to account for the observed
1161.8~\AA\ emission of 0.24 rayleighs is 2\%.  This is consistent
with the relative population of this level seen in the laboratory
experiments \citep{Zhang:2005}, when we consider that these experiments
are done at discrete wavelengths while in the cometary context
photodissociation is produced by the wavelength integrated solar flux.

We conclude that photodissociation of \Htwoco\ is sufficient to produce
vibrationally excited \Htwo\ that accounts for the presence of
Lyman-$\alpha$ excited fluorescence of \Htwo\ in the coma.  While the
production rate of \Htwoco\ is $\sim$0.5\% relative to \Htwoo, the
solar photodissociation rate into \Htwo\ is
$\sim$100 times higher for \Htwoco, depending on the solar activity
level \citep{Huebner:1992}.  Thus both molecules contribute comparable
amounts of \Htwo\ to the coma.  The branching of water photodissociation
into \Htwo\ + O($^1$D) has also been studied in the laboratory
\citep{Slanger:1982}, but there are no available measurements of the
resultant \Htwo\ level populations.  Although there has been extensive
theoretical work on the H + OH channel of \Htwoo\ photodissociation
\citep{Schinke:1993}, there does not seem any devoted to the \Htwo\ + O($^1$D)
channel, which is clearly warranted.  On the basis of the brightness
of the Lyman-$\beta$ pumped fluorescent lines, \citet{Feldman:2002}
suggested that solar photodissociation of water led to a thermal population
of \Htwo\ molecules in the $v = 0$, $J = 1$ level.

\subsection{Caveat on Temporal Variability \label{caveat}}

\placefigure{a2_temporal}

We noted above the utility of the SOHO/SWAN data of \citet{Combi:2008}
to obtain daily values of the water production rate.  For C/2001~A2
near the time of the \fuse\ observations, the comet exhibited several
visual outbursts, particularly one on 2001~July~11.5, one day before
the \fuse\ observations, which saw a 1.5~magnitude increase in visual
brightness \citep{Sekanina:2002}.  These outbursts were reflected in
sudden increases in $Q_{\rm H_2O}$ of factors of 2--3 in one day in the
data of \citeauthor{Combi:2008}.  The \fuse\ observations of
this comet spanned 7.5 hours during which the CO emission decreased
by a factor of two (Fig.~\ref{a2_temporal}), likely a manifestation of
the decrease in coma abundance following the outburst.  Note that for
this comet the \fuse\ $30'' \times 30''$ aperture projects to $6500
\times 6500$~km on the sky so that a molecule moving radially outward
at 1~\kms\ would exit the field-of-view in less than one hour.  Hence
the observed variation represents variability of the CO source on this
time scale.  On the other hand, the column densities given in
Table~\ref{tabn} are taken from the summed spectra (to improve the
signal/noise ratio) and thus represent average values.  Caution is
therefore urged in deriving relative production rates from diverse
observations, even if obtained on the same date, and in the comparison
of results from different observers.  

The July~11.5 outburst makes it difficult to compare the derived
\Htwoco\ production rates of \citet{Gibb:2007} on July~9.5 and 10.5,
noted above, with the present results.  From the limited data
available, it is not possible to determine whether \Htwoco\ production
was enhanced by the outburst or whether the increase inferred from the
\fuse\ spectra is just the result of intrinsic variability responsible
for the change seen by \citeauthor{Gibb:2007} a few days earlier.
The same holds for a comparison of our results with the CO production
rate derived from the same infrared spectrum on July~10.5
\citep{Magee:2008} even though in this case the values of $Q_{\rm CO}$
are comparable.

\placefigure{q4_temporal}

A similar caveat applies to comet C/2001~Q4 which \fuse\ observed over
a 27~h period.  The variation of the CO emission is shown in
Fig.~\ref{q4_temporal} and can be fit with a sinusoidal curve of period
17.0~h and 22\% amplitude.  Yet the CO production rate, derived
from \hst/STIS observations of the CO Fourth Positive system
\citep{Lupu:2007}, made near a minimum in the \fuse\ lightcurve, is
consistent with the mean value derived from the \fuse\ spectra.  In
contrast, comet C/2001~WM1 showed little variation in CO emission over
a 60~h period.  For this comet, simultaneous \hst/STIS observations
were also obtained and \citeauthor{Lupu:2007} derived comparable CO
production rates from both sets of spectra.

\subsection{Relative Abundance of \cotwo\ \label{co2}}

As noted in Section~\ref{electron}, for molecules with long lifetimes
against solar photodissociation such as \cotwo, electron impact
dissociation becomes an important destruction channel.  Using the
electron impact rates given in Table~\ref{tabg}, however, would still require an
abnormally large ($\geq 20$\%) abundance of \cotwo\ to account for the derived
``hot'' CO column density.  Since there are no other suitable parents
for CO other than \cotwo\ and \Htwoco, it is likely that the model used
to calculate the values given in Table~\ref{tabg} significantly
underestimates the electron impact contribution to
\cotwo\ dissociation.  \citet{Cosby:1992} note that the threshold for
this process is at 7.4~eV whereas the channel leading to the
\atpi\ state of CO requires 11.4~eV.  At low electron energies, the
model of \citet{Koros:1987} does not match the data of
\citet{Gringauz:1986}, which appears to follow an $E^{-1}$
distribution, so that there may be an additional large contribution to
the dissociation rate from low energy electrons.  There are also
uncertainties in the radial and angular distribution of the electrons
that make our estimate quantitatively unreliable.  A more complete
analysis of cometary photoelectrons, which is beyond the scope of the
present paper, would entail an analysis of CO Cameron band emission
similar to that presented by \citet{Weaver:1994}.  Fortunately,
relevant \hst/STIS spectra do exist for comet C/2001~Q4 and will be
treated in a separate publication.

\subsection{Distributed Source of CO \label{dist}}

The first evidence for a distributed source of CO in the coma of a
comet came from the {\it in situ} Neutral Mass Spectrometer experiment
on the {\it Giotto} mission to comet 1P/Halley
\citep{Eberhardt:1987,Eberhardt:1999}.  \citet{Meier:1993} also showed,
from the same data set, that the \Htwoco\ also appeared to come from an
extended source and that the \Htwoco\ was sufficient to account for the
extended CO \citep{Eberhardt:1999}.  Polyoxymethylene (POM), evaporated
from organic material on refractory grains, has been postulated as the
source of this \Htwoco\ \citep{Cottin:2004}.  In recent years, as high
resolution spectroscopic instrumentation in the near infrared has made
the observation of CO in comets almost routine, considerable effort has
been expended in trying to determine if such distributed sources are
common or unique to certain comets \citep{Bockelee-Morvan:2004}.
Identification of a distributed emission source is based on long-slit
spectroscopy in which the spatial profile of the emission deviates from
that expected for radial outflow \citep{DiSanti:2001,Brooke:2003}.  
For CO this can be done in both the infrared and the ultraviolet.

This approach is applicable only to the brightest comets because the
spatial pixels along the slit must be small enough to provide adequate
spatial resolution.  For such comets, such as C/1995 O1 (Hale-Bopp),
the CO column densities along the line-of-sight (to both the Sun and
to the observer) are sufficiently large that optical depth effects
tend to produce a flattening of the spatial profile near the nucleus
that mimics the effect of a distributed source \citep[the ultraviolet
case has been discussed by][]{Lupu:2007}.  This concern was
addressed by \citeauthor{DiSanti:2001} in their analysis of CO
in comet Hale-Bopp.  They found a very high production rate of CO
relative to that of water, 24.1\%, with half being from the nucleus
and the other half from a distributed source.  In the same comet,
\citet{Brooke:2003} found an even higher relative abundance of CO,
37--41\%, with 90\% coming from a distributed source.
\citet{Bockelee-Morvan:2005} questioned both of these results
suggesting that they were the effect of a large infrared opacity,
even though both \citeauthor{DiSanti:2001} and \citeauthor{Brooke:2003}
accounted for opacity effects.  In
any case, if such a distributed source were \Htwoco\ it would imply an
abundance in this comet an order of magnitude higher than that cited by
\citet{Biver:2006}.  

\section{CONCLUSION}

Spectroscopic observations of four comets by \fuse\ have opened a new
window for the investigation of the physics and chemistry of the
cometary coma.  The high spectral resolution of \fuse\ has made it
possible to separate the multiple components of the CO
emission, while the high sensitivity has allowed the detection of
fluorescent \Htwo\ emission pumped by solar \ion{O}{6} radiation.
Both of these can be understood in terms of the dissociation processes
of \Htwoco\ and \cotwo.  Despite the small sample, it is clear that
previous ultraviolet spectroscopic observations of CO in comets
including those from sounding rockets, \iue\ and \hst\ \citep[e.g.,
][]{Feldman:1997} probably included these additional components of CO
that were not resolved, either spectrally or spatially, from the native
(``cold'') CO with the instrumentation used.  Photoelectron excitation
contributes $\sim$10\% of the observed emission of the CO $C-X$ (0,0)
band and needs to be accounted for in deriving CO production rates from
\fuse\ observations.

\acknowledgments

We thank the \fuse\ ground system personnel, particularly B.\ Roberts,
T.\ Ake, A.\ Berman, B.\ Gawne, and J.\ Andersen, for their efforts in
planning and executing these moving target observations.  This work is
partially based on data obtained for the Guaranteed Time Team by the
NASA-CNES-CSA \fuse\ mission operated by the Johns Hopkins University.
Financial support was provided by NASA contract NAS5-32985 and NASA
grant NAG5-12963.

{\it Facilities:} \facility{FUSE ()}

%%%%%%%%%%%%%  REFERENCES  %%%%%%%%%%%%%%%%%%%%%%%%%%%%%%%%%%%%%

\renewcommand\baselinestretch{1.2}%

\begin{table}
\begin{center}
\caption{Comet observation parameters and observed CO brightnesses. \label{tab1}} 
\medskip
\begin{tabular}{@{}lcccc@{}}
\tableline\tableline
Comet & C/1999 T1 & C/2001 A2 & C/2000 WM1 & C/2001 Q4 \\
 & (McNaught-Hartley) & (LINEAR) & (LINEAR) & (NEAT)\\
\tableline
Date  & 2001 Feb 3 & 2001 Jul 12 & 2001 Dec 7--10 & 2004 Apr 24 \\
Start Time (UT) & 17:43 & 13:48 & 08:48 & 00:40 \\
Exposure Duration (s) & 15,614 & 15,063 & 36,467 & 68,282 \\
$r$ (AU) &  1.43 & 1.20 & 1.12 & 1.03 \\
$\Delta$ (AU) & 1.29 & 0.30 & 0.34 & 0.51 \\
$\dot{r}$ (km s$^{-1}$) & 14.9 & 22.8 & --28.3 & --10.8 \\
$F_{10.7}$ (sfu) & 160 & 140 & 220 & 115 \\
\tableline
$C-X$~(0,0) (rayleighs) & $1.46 \pm 0.03$ & $2.47 \pm 0.04$ & $0.74 \pm 0.02$ & $6.29 \pm 0.03$ \\
$\ldots$P(40) peak (rayleighs) & $\leq 0.02$ & $0.16 \pm 0.03$ & $0.10 \pm 0.01$ & $0.16 \pm 0.02$ \\
$B-X$~(0,0) (rayleighs) & $0.19 \pm 0.02$ & $0.71 \pm 0.03$ & $0.16 \pm 0.01$ & $1.51 \pm 0.02$ \\
\tableline
\end{tabular}
\end{center}
\end{table}

%\input{h2co_tablen_rev}

%revised 3/24/2009
\begin{table}
\begin{center}
\caption{Derived CO column densities and rotational temperatures in the
\fuse\ $30'' \times 30''$ aperture.  Errors are from the 1$\sigma$
contours in a three parameter fit. \label{tabn}}
\medskip
\begin{tabular}{@{}lcccc@{}}
\tableline\tableline
Comet & C/1999 T1 & C/2001 A2 & C/2000 WM1 & C/2001 Q4 \\
 & (McNaught-Hartley) & (LINEAR) & (LINEAR) & (NEAT)\\
\tableline
$N_{\rm cold}$ (cm$^{-2}$) & $(3.5\pm0.2) \times 10^{13}$ & $(2.7\pm0.15) \times 10^{13}$ & $(6.0\pm0.4) \times 10^{12}$ & $(1.55\pm0.05) \times 10^{14}$ \\
$N_{\rm hot}$ (cm$^{-2}$)  & $\leq1.3 \times 10^{12}$ & $(1.10\pm0.06) \times 10^{13}$ & $(2.5\pm0.5) \times 10^{12}$ & $(2.0\pm0.1) \times 10^{13}$ \\
$N_{\rm non-thermal}$ (cm$^{-2}$)\tablenotemark{a}  & $\leq0.8 \times 10^{12}$ & $(4.5\pm0.7) \times 10^{12}$ & $(2.5\pm0.3) \times 10^{12}$ & $(3.3\pm0.3) \times 10^{12}$ \\
$T_{\rm cold}$ (K) & $59\pm5$ & $56\pm7$ & $75\pm10$ & $58\pm2$ \\ 
$T_{\rm hot}$ (K) & 500 & 500 & 500 & 600 \\ 
\tableline 
\end{tabular}
\end{center}
\tablenotetext{a}{Assumes an equal contribution from the R(40) peak not
observed.  Uncertainties are 1$\sigma$ in the counting statistics.}
\end{table}

%\input{h2co_tableg_rev}
%revised 3/24/2009
\begin{table}
\begin{center}
\caption{Derived effective electron impact excitation rates ($g'$), in
molecule$^{-1}$~s$^{-1}$, averaged over the \fuse\ aperture. \label{tabg}}
\medskip
\begin{tabular}{@{}lcccc@{}}
\tableline\tableline
Comet & C/1999 T1 & C/2001 A2 & C/2000 WM1 & C/2001 Q4 \\
 & (McNaught-Hartley) & (LINEAR) & (LINEAR) & (NEAT)\\
\tableline
$B-X$~(0,0)\tablenotemark{a} & $9.0 \times 10^{-9}$ & $2.4 \times 10^{-8}$ & $2.1 \times 10^{-8}$ & $7.4 \times 10^{-9}$ \\
$C-X$~(0,0)\tablenotemark{a} & $5.8 \times 10^{-9}$ & $1.6 \times 10^{-8}$ & $1.3 \times 10^{-8}$ & $4.7 \times 10^{-9}$ \\
$a-X$~(all bands)\tablenotemark{b} & $2.6 \times 10^{-7}$ & $7.1 \times 10^{-7}$ & $6.1 \times 10^{-7}$ & $2.2 \times 10^{-7}$ \\
total dissociation\tablenotemark{b} & $1.1 \times 10^{-6}$ & $2.9 \times 10^{-6}$ & $2.4 \times 10^{-6}$ & $8.6 \times 10^{-7}$ \\ 
\tableline 
\end{tabular}
\end{center}
\tablenotetext{a}{electron impact on CO at 1 AU.  For comparison, the
optically thin fluorescence efficiencies for the $B-X$~(0,0) and
$C-X$~(0,0) bands at 1 AU are $2.3 \times 10^{-9}$ and $1.0 \times 10^{-7}$
molecule$^{-1}$~s$^{-1}$, respectively, for solar maximum conditions.}
\tablenotetext{b}{electron impact on CO$_2$ at 1 AU.}
\end{table}

%\input{h2co_tableq_rev}

%revised 3/24/2009
\begin{table}
\begin{center}
\caption{Modeled production rates and production rate ratios. \label{tabq}} 
\medskip
\begin{tabular}{@{}lcccc@{}}
\tableline\tableline
Comet & C/1999 T1 & C/2001 A2 & C/2000 WM1 & C/2001 Q4 \\
 & (McNaught-Hartley) & (LINEAR) & (LINEAR) & (NEAT)\\
\tableline
$Q_{\rm CO}$ (s$^{-1}$)\tablenotemark{a} & $7.8 \times 10^{27}$ & $1.33 \times 10^{27}$ & $3.7 \times 10^{26}$ & $1.45 \times 10^{28}$ \\
$Q_{\rm H_2CO}$ (s$^{-1}$) & $\leq 3.7 \times 10^{26}$ & $7.6 \times 10^{26}$ & $4.6 \times 10^{26}$ & $8.1 \times 10^{26}$ \\
\tableline
$Q_{\rm H_2O}$ (s$^{-1}$) & $7.6 \times 10^{28}$ & $6.0 \times 10^{28}$ & $8.8 \times 10^{28}$ & $2.0 \times 10^{29}$ \\ 
\tableline
$Q_{\rm CO}$/$Q_{\rm H_2O}$ (\%)\tablenotemark{a} & 10.3 & 2.2 & 0.41 & 7.2 \\ 
$Q_{\rm H_2CO}$/$Q_{\rm H_2O}$ (\%) & $\leq$0.5 & 1.3 & 0.52 & 0.40 \\ 
\tableline 
\end{tabular}
\end{center}
\tablenotetext{a}{corrected for photoelectron impact contribution.}
\end{table}

\renewcommand\baselinestretch{1.6}%

%\newpage
\clearpage 
\begin{center}{\bf FIGURE CAPTIONS}\end{center}

\figcaption[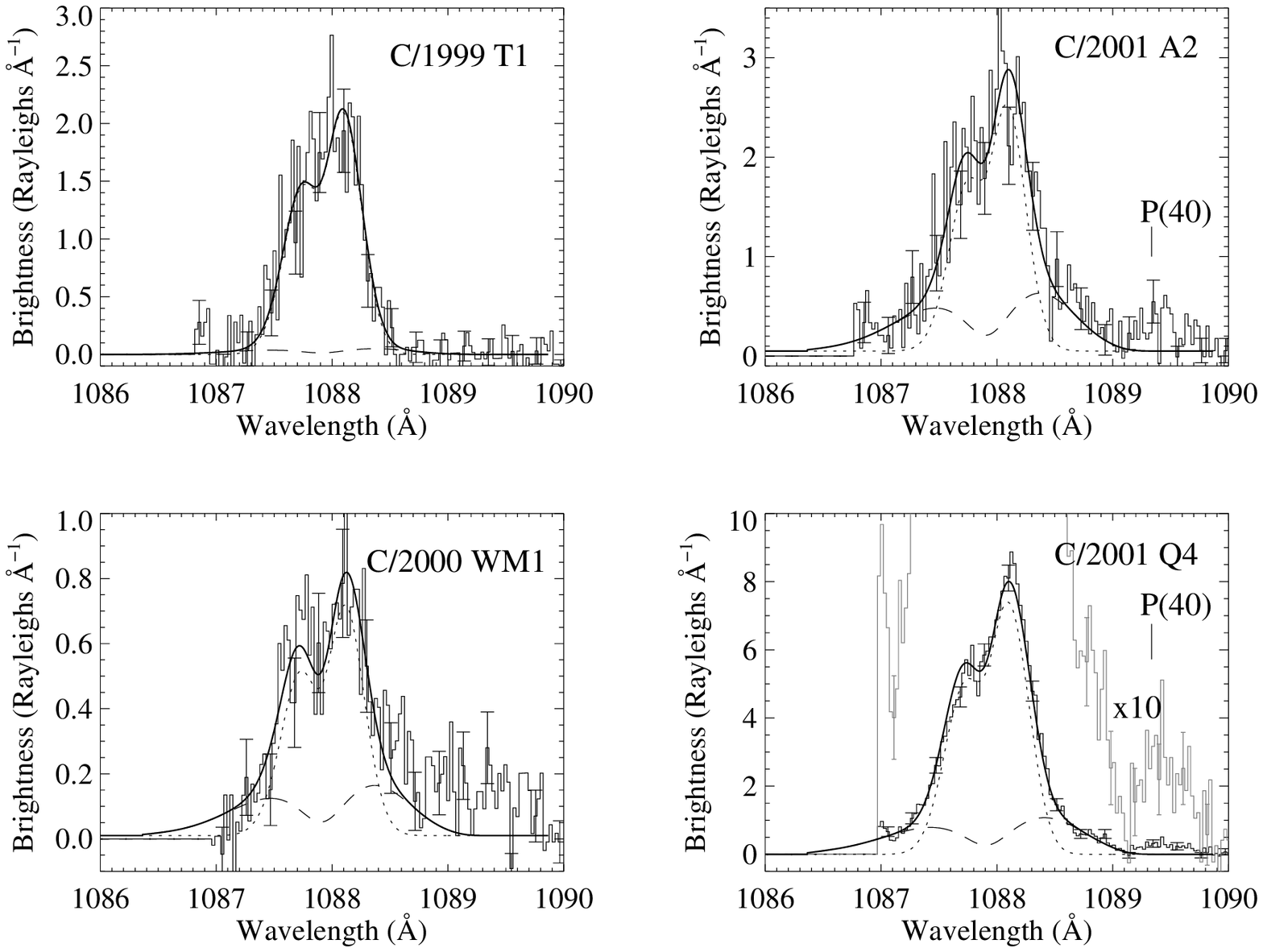]{CO $C-X$ (0,0) band in comets McNaught-Hartley, 
C/2001~A2, C/2000~WM1 and C/2001~Q4.  The model fit is shown together with the
``cold'' (dotted) and ``hot'' (dashed) components. P(40) indicates the
position of the P-branch transitions originating from $J$ levels
near the peak of the hot, non-thermal rotational distribution seen
in laboratory photodissociation experiments \citep{vanZee:1993}. The
corresponding R-branch lines lie below the short wavelength cut-off of
the LiF2a detector.  For C/2001~Q4, the gray line represents the
vertical scale expanded by a factor of 10 to enhance the visibility
of the non-thermal peak.  \label{cx1}}

\figcaption[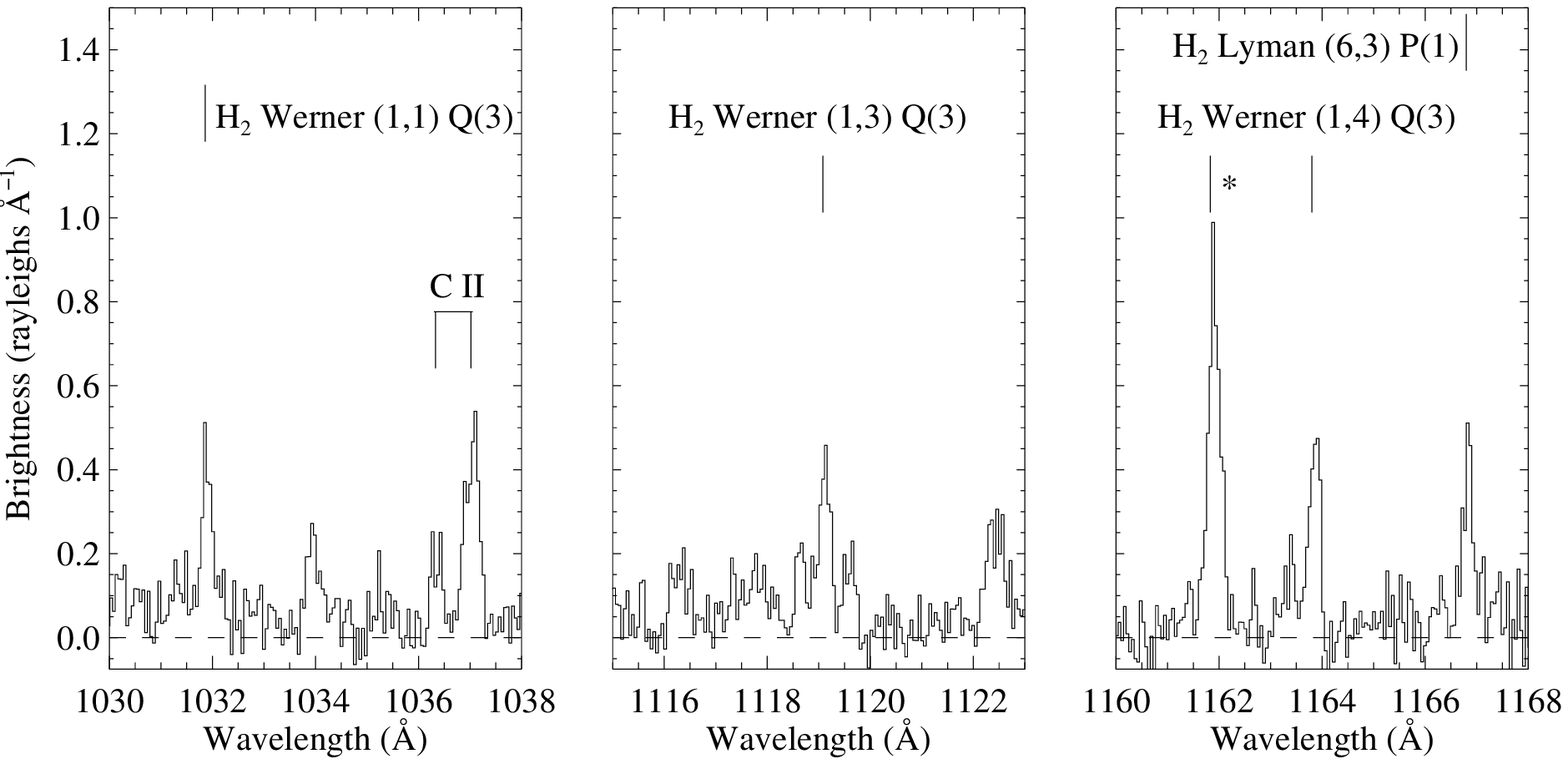]{\ion{O}{6} pumped \Htwo\ lines in the \fuse\ 
spectrum of comet C/2001 Q4.  The wavelength scale is in the comet's
rest frame.  The exposure time for the LiF1a channel (left panel) was
67,360~s, while the other two panels include the sum of LiF1b and LiF2a
with a total exposure time of 137,530~s.  The feature indicated by (*)
is identified as solar Lyman-$\alpha$ pumped fluorescence of \Htwo\ in
the Lyman (1,1)P(5) and R(6) lines.  The Lyman (6,3)P(1) line is pumped
by solar Lyman-$\beta$.  \label{h2}}

\figcaption[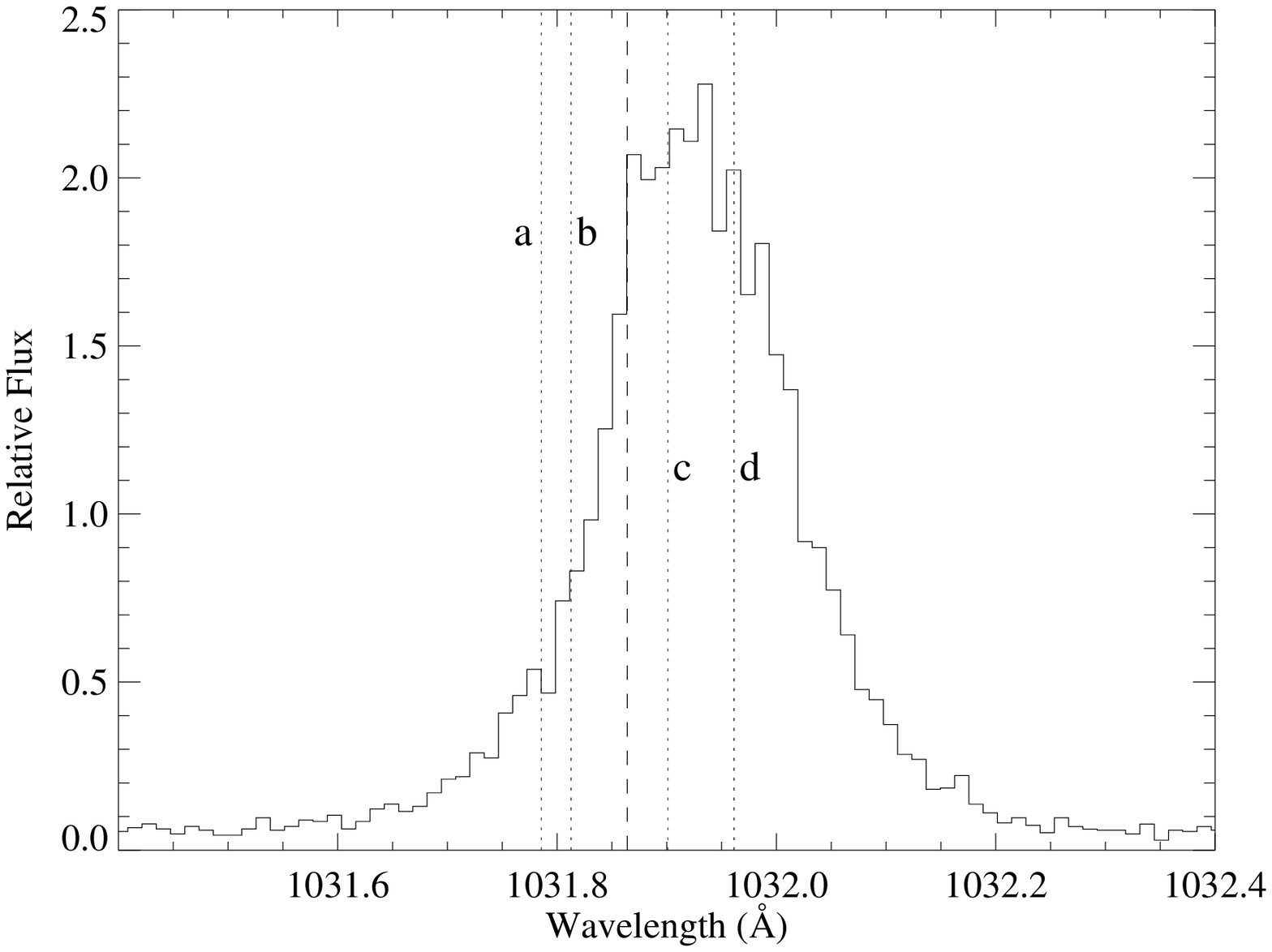]{Solar \ion{O}{6} line profile derived from an instrumentally scattered \fuse\ spectrum.  The position of the Werner
(1,1)~Q(3) line is shown as the dashed line.  The heliocentric Doppler
shifts of the four comets studied are indicated as (a) C/2001 A2, (b)
McNaught-Hartley, (c) C/2001 Q4, (d) C/2000 WM1.  \label{solar}}

\figcaption[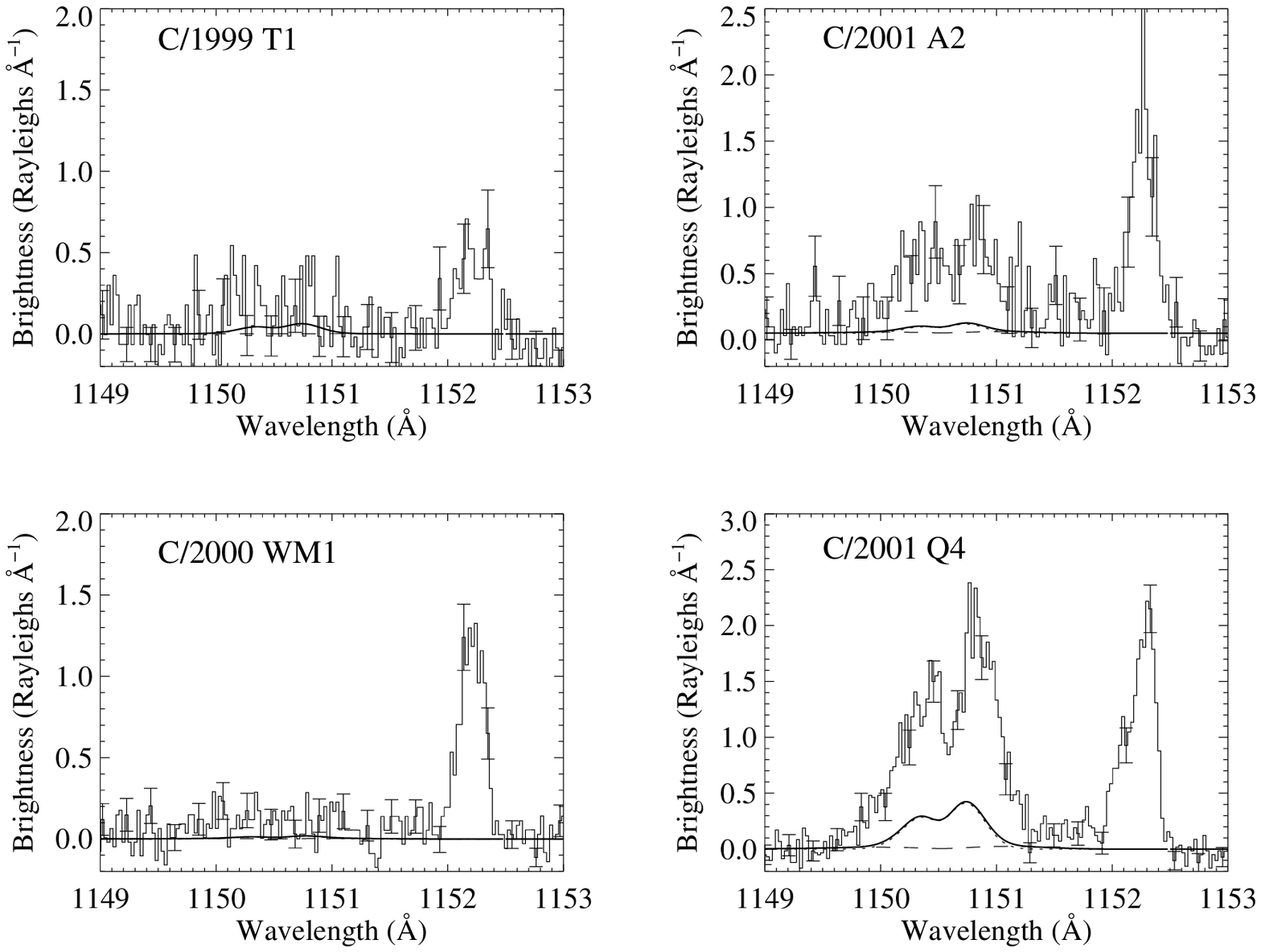]{Same as Fig.~\ref{cx1} for the CO $B-X$ (0,0) band.  The predicted solar fluorescence brightness is shown for both
the ``cold'' (dotted) and ``hot'' (dashed) components.  The shape of
the neighboring \ion{O}{1}~$\lambda$1152.15 feature reflects the
spatial distribution of the emitter in the \fuse\ aperture.
\label{bx1}}

\figcaption[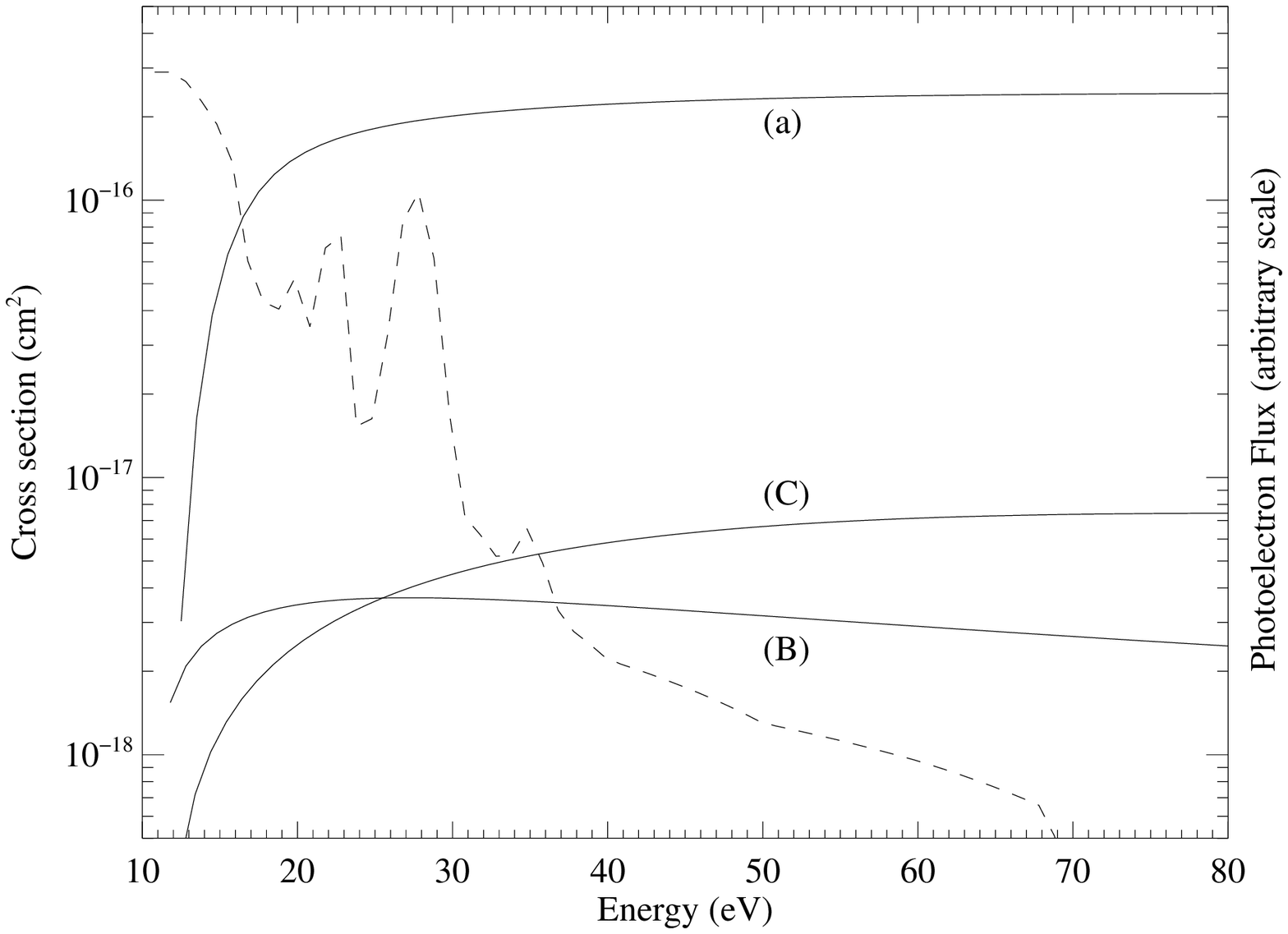]{Cross sections for electron impact excitation  
\citep[from][]{Shirai:2001}.  The curves labeled (B) and (C) are for
excitation of CO to produce $B-X$~(0,0) and $C-X$~(0,0) band photons,
respectively, while (a) represents electron impact dissociation of
\cotwo\ into the \atpi\ state of CO.  The dashed curve represents a
theoretical photoelectron energy distribution \citep[from][]{Koros:1987}.
\label{cross}}

\figcaption[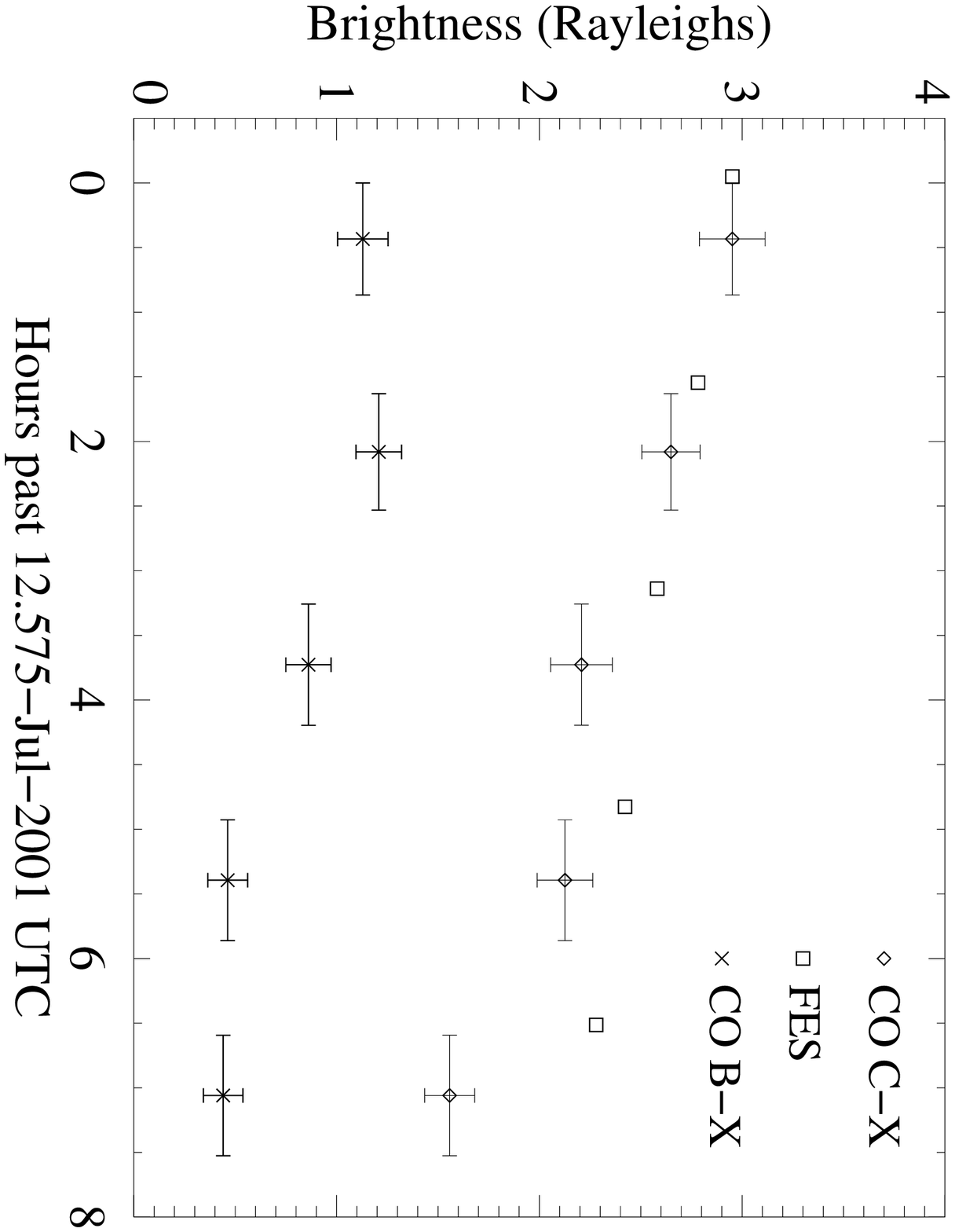]{Temporal variability of the CO $C-X$ and $B-X$ 
bands in comet C/2001~A2.  The horizontal bars indicate the individual
exposures.  The visual brightness deduced from the \fuse\ Fine Error
Sensor is also shown.  \label{a2_temporal}}

\figcaption[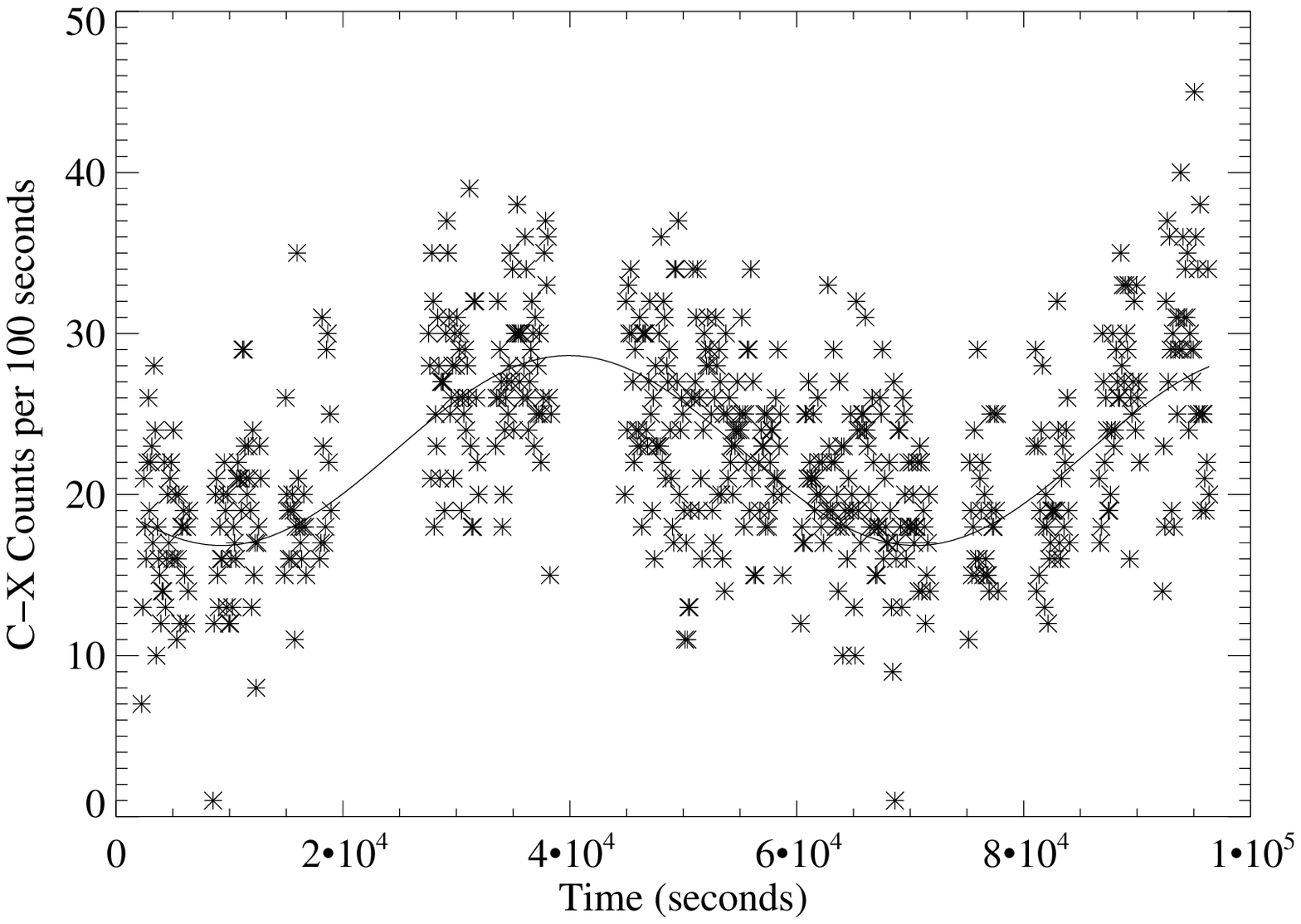]{Temporal variability of the CO $C-X$ band in comet
C/2001~Q4.  The origin of time is UT 2004~April~24~00:40:32.  The solid
line is a sinusoidal fit with a period of 17.0~h and an amplitude of
22\%.  \label{q4_temporal}}

\renewcommand\baselinestretch{1.2}%
\setcounter{figure}{0}

\begin{figure}[ht]
\begin{center}
\epsscale{1.0}
\rotatebox{0.}{
\plotone{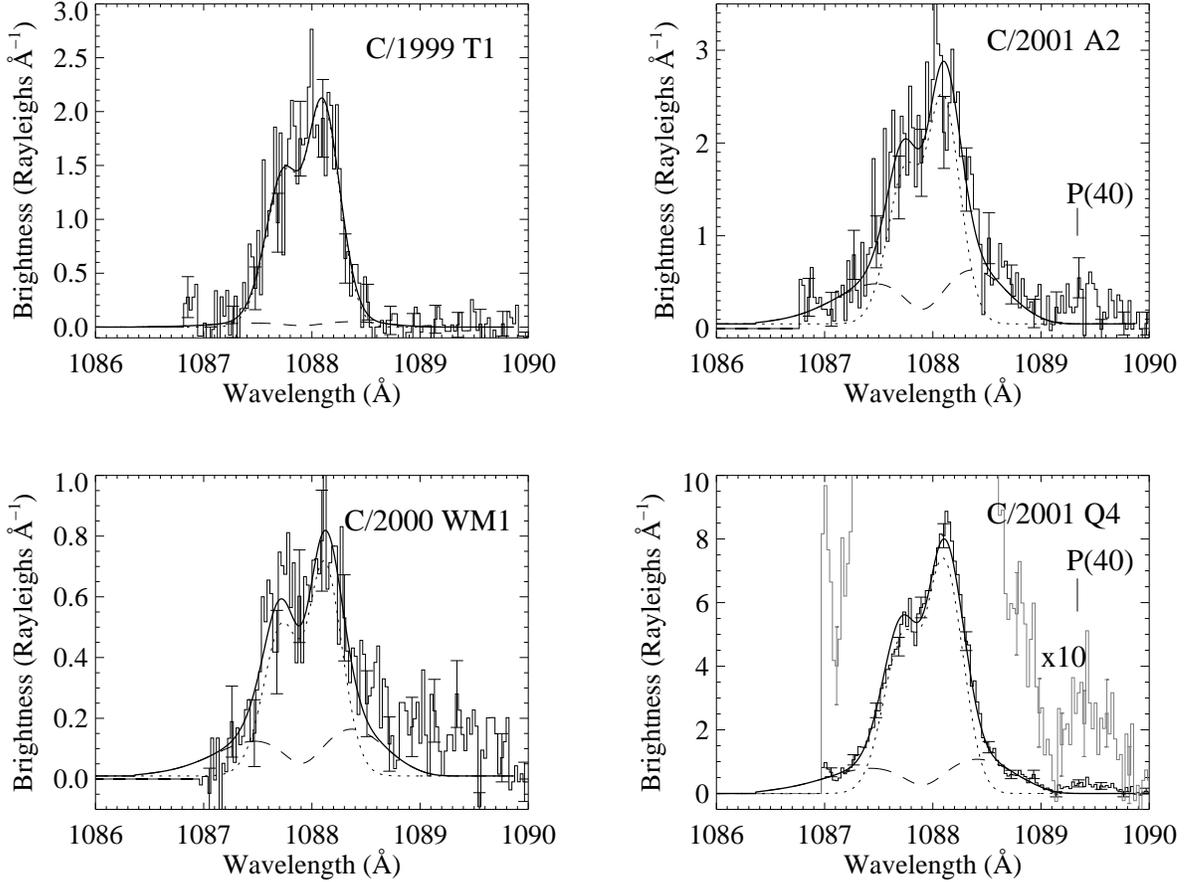}} 
%\vspace{-0.45in}
\caption {CO $C-X$ (0,0) band in comets McNaught-Hartley, 
C/2001~A2, C/2000~WM1 and C/2001~Q4.  The model fit is shown together with the
``cold'' (dotted) and ``hot'' (dashed) components. P(40) indicates the
position of the P-branch transitions originating from $J$ levels
near the peak of the hot, non-thermal rotational distribution seen
in laboratory photodissociation experiments \citep{vanZee:1993}. The
corresponding R-branch lines lie below the short wavelength cut-off of
the LiF2a detector.  For C/2001~Q4, the gray line represents the
vertical scale expanded by a factor of 10 to enhance the visibility
of the non-thermal peak.  %\label{cx1}}
}
\end{center}
\end{figure}

\begin{figure}[ht]
\begin{center}
\epsscale{1.0}
\rotatebox{0.}{
\plotone{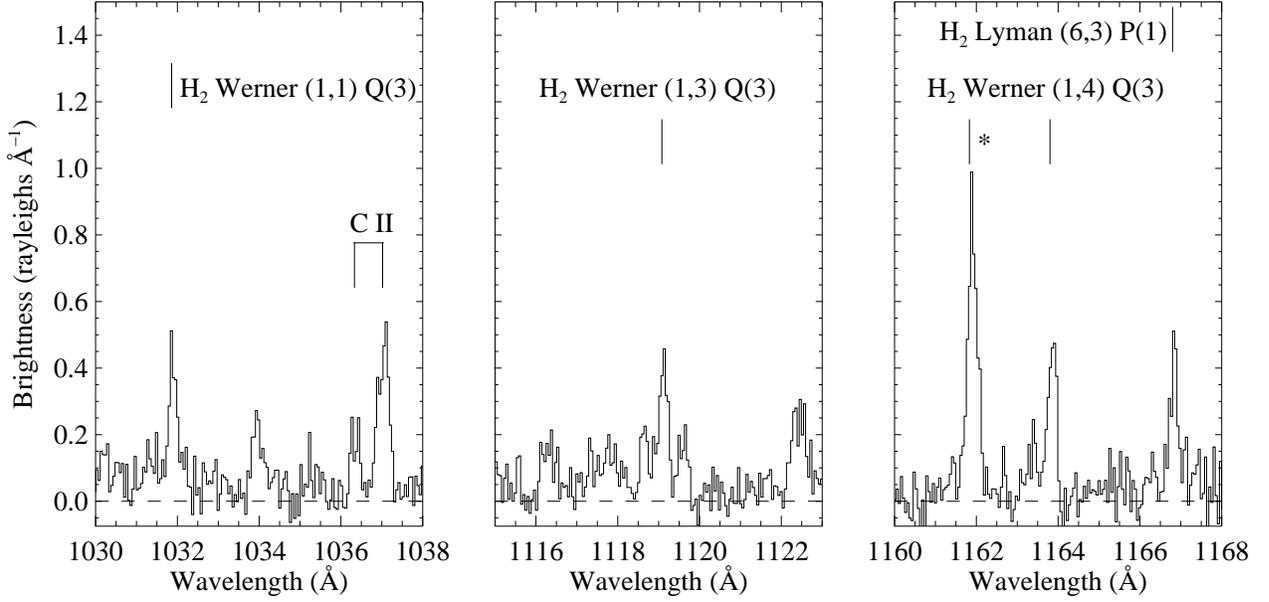}} 
%\vspace{-0.45in}
\caption {\ion{O}{6} pumped \Htwo\ lines in the \fuse\ spectrum of
comet C/2001 Q4.  The wavelength scale is in the comet's rest frame.
The exposure time for the LiF1a channel (left panel) was 67,360~s, while
the other two panels include the sum of LiF1b and LiF2a with a total
exposure time of 137,530~s.  The feature indicated by (*) is identified
as solar Lyman-$\alpha$ pumped fluorescence of \Htwo\ in the Lyman 
(1,1)P(5) and R(6) lines.  The Lyman (6,3)P(1) line is pumped by
solar Lyman-$\beta$.  %\label{h2}}
}
\end{center}
\end{figure}

\begin{figure}[ht]
\begin{center}
\epsscale{1.0}
\rotatebox{0.}{
\plotone{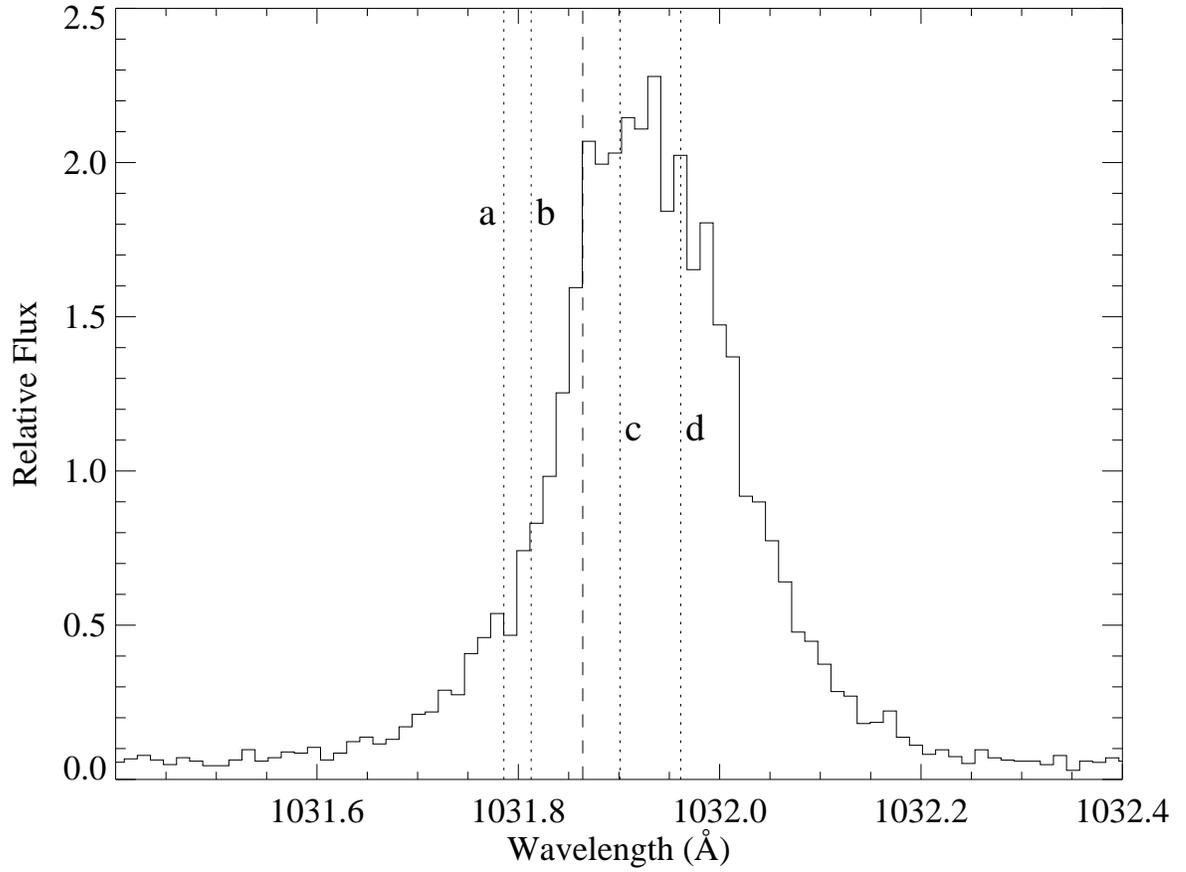}} 
%\vspace{-0.45in}
\caption {Solar \ion{O}{6} line profile derived from an instrumentally
scattered \fuse\ spectrum.  The position of the Werner (1,1)~Q(3) line
is shown as the dashed line.  The heliocentric Doppler shifts of the
four comets studied are indicated as (a) C/2001 A2, (b)
McNaught-Hartley, (c) C/2001 Q4, (d) C/2000 WM1.  %\label{solar}}
}
\end{center}
\end{figure}

\begin{figure}[ht]
\begin{center}
\epsscale{1.0}
\rotatebox{0.}{
\plotone{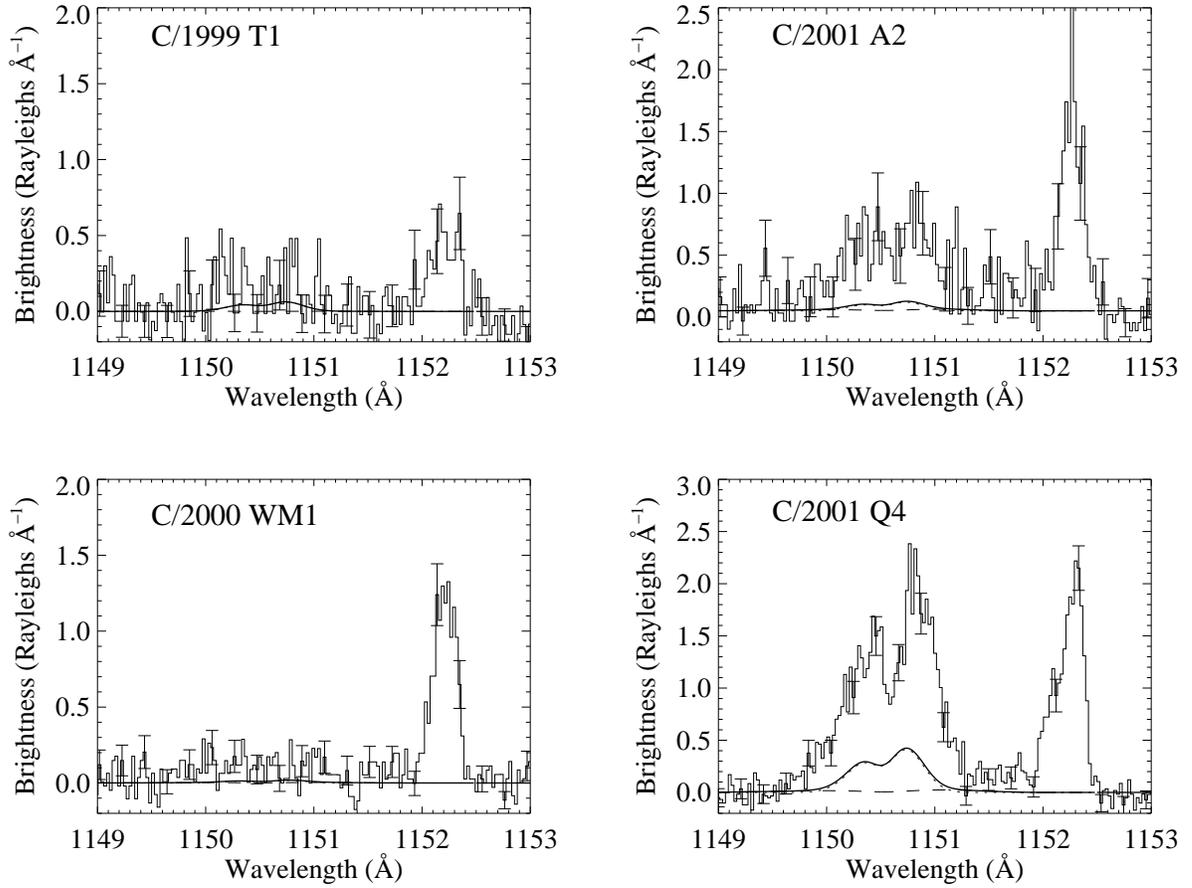}} 
\caption {Same as Fig.~\ref{cx1} for the CO $B-X$ (0,0) band.  The 
predicted solar fluorescence brightness is shown for both the
``cold'' (dotted) and ``hot'' (dashed) components.  The shape of the
neighboring \ion{O}{1}~$\lambda$1152.15 feature reflects the spatial
distribution of the emitter in the \fuse\ aperture.  %\label{bx1}}
}
\end{center}
\end{figure}

\begin{figure}[ht]
\begin{center}
\epsscale{1.0}
\rotatebox{0.}{
\plotone{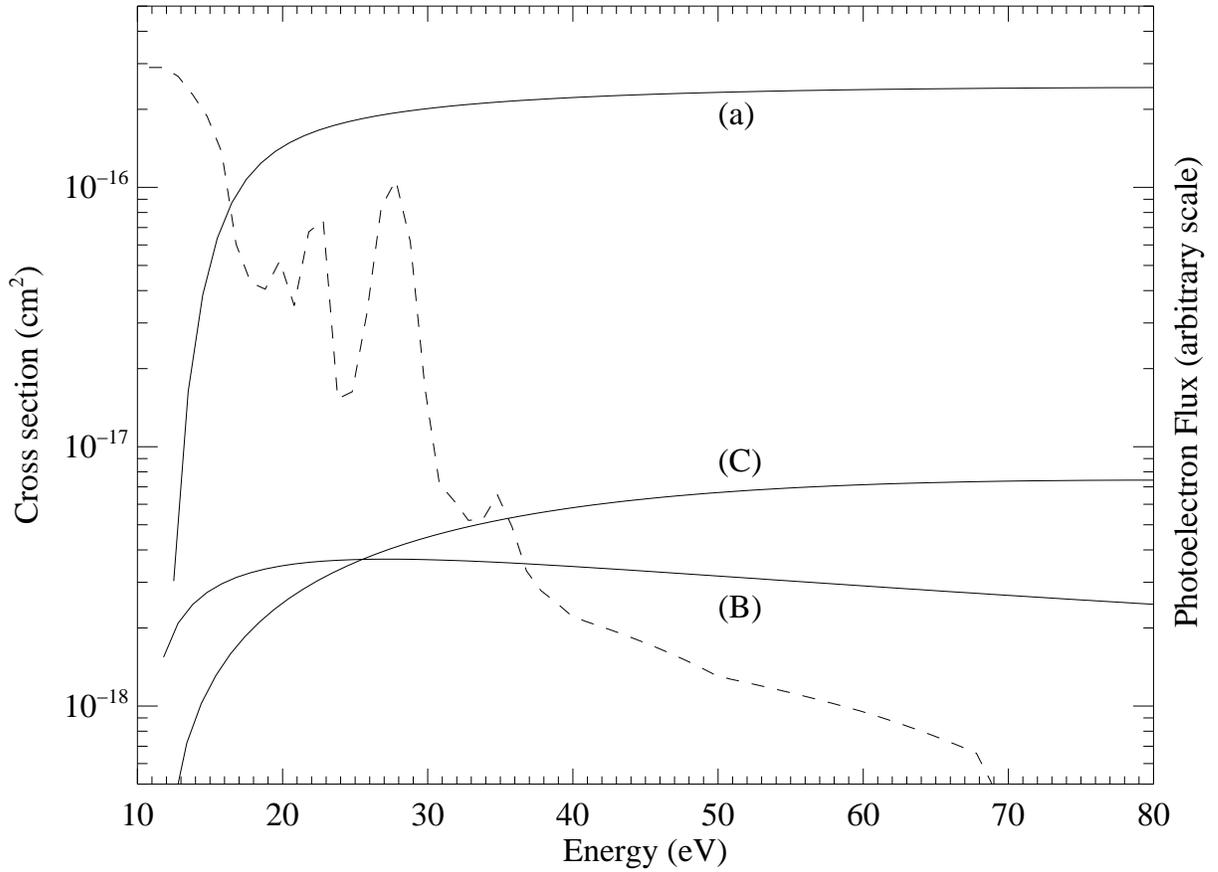}} 
\caption {Cross sections for electron impact excitation  
\citep[from][]{Shirai:2001}.  The curves labeled (B) and (C) are for
excitation of CO to produce $B-X$~(0,0) and $C-X$~(0,0) band photons,
respectively, while (a) represents electron impact dissociation of
\cotwo\ into the \atpi\ state of CO.  The dashed curve represents a
theoretical photoelectron energy distribution \citep[from][]{Koros:1987}.
%\label{cross}}
}
\end{center}
\end{figure}

\begin{figure}[ht]
\begin{center}
\epsscale{0.7}
\rotatebox{90.}{
\plotone{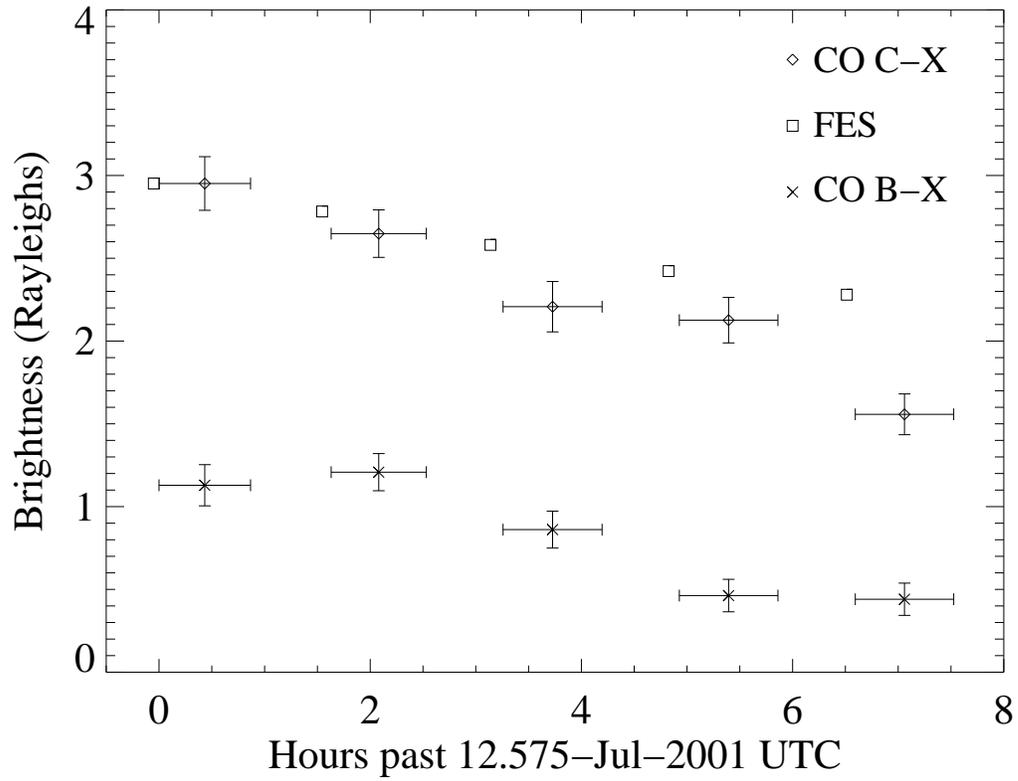}} 
\caption {Temporal variability of the CO $C-X$ and $B-X$ bands in comet
C/2001~A2.  The horizontal bars indicate the individual exposures.  The
visual brightness deduced from the \fuse\ Fine Error Sensor is also
shown.  %\label{a2_temporal}}
}
\end{center}
\end{figure}

\begin{figure}[ht]
\begin{center}
\epsscale{1.0}
\rotatebox{0.}{
\plotone{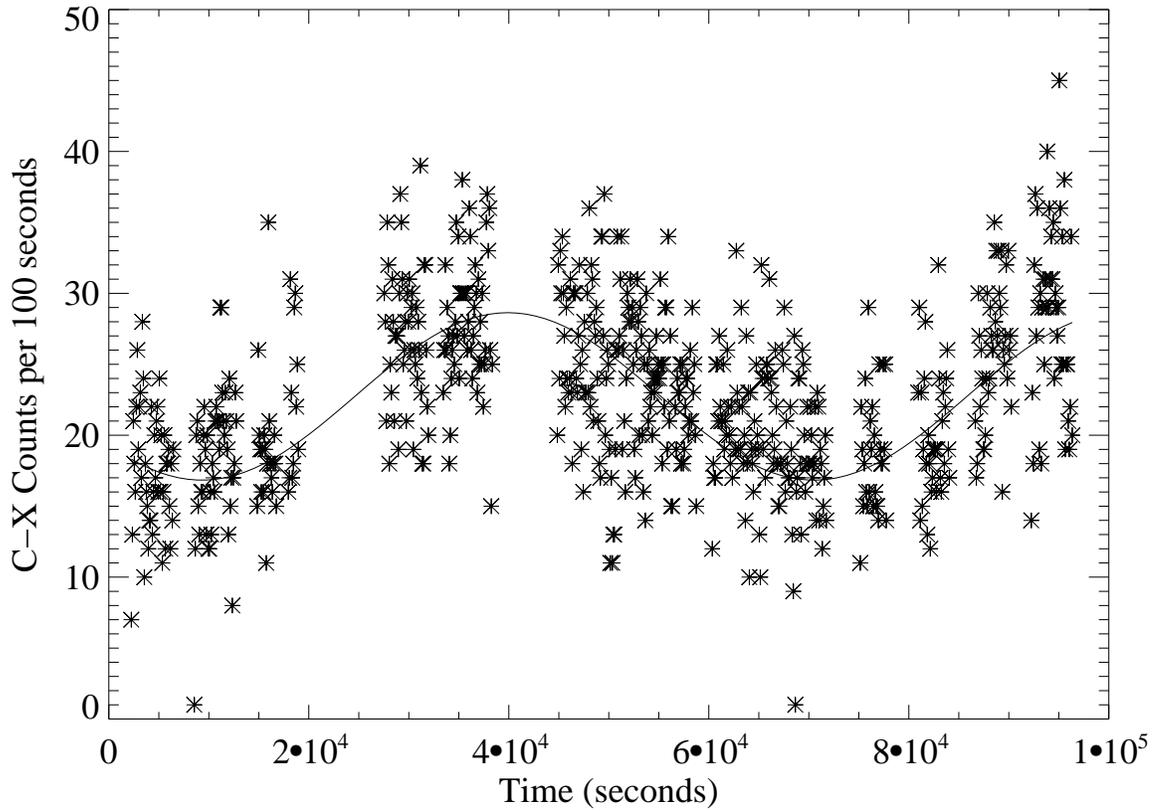}} 
%\plotone{cxperiod.eps}} 
\caption {Temporal variability of the CO $C-X$ band in comet
C/2001~Q4.  The origin of time is UT 2004~April~24~00:40:32.  The solid
line is a sinusoidal fit with a period of 17.0~h and an amplitude of
22\%.  %\label{q4_temporal}}
}
\end{center}
\end{figure}

\renewcommand\baselinestretch{1.6}%

\end{document}